\documentclass[preprint,prd,showpacs,amsmath,amssymb,amsthm,nofootinbib]{revtex4}
\usepackage[mathscr]{euscript}
\usepackage{bm}
\usepackage{graphicx}
\usepackage{subfigure}
\usepackage{overpic}
\usepackage{multirow}
\usepackage{booktabs}
\usepackage{array}
\usepackage{floatrow}
\usepackage{float} 
\usepackage{amsmath,amssymb,amsthm}
\usepackage{cases}
\usepackage[colorlinks=true,linkcolor=red]{hyperref}
\newfloatcommand{capbtabbox}{table}[][\FBwidth]
\newcommand{\bea}{\begin{eqnarray}}
\newcommand{\eea}{\end{eqnarray}}
\newcommand{\beq}{\begin{equation}}
\newcommand{\eeq}{\end{equation}}

\def\/{\over}
\usepackage{ulem}

\begin{document}

\title{Reconciling the ACT  Preference in $f(T)$  Gravity: Inflation and Reheating Constraints}

\author{ Feng-Yi Zhang$^{1,2}$\footnote{zfy@usc.edu.cn},   Rongrong Zhai$^{3}$\footnote{rrzhai@xztu.edu.cn} and Li-Yang Chen$^{4}$\footnote{Corresponding author: lychen@cdnu.edu.cn} }
\affiliation{$^1$School of Mathematics and Physics, University of South China, Hengyang, 421001, China\\
$^2$Hunan Key Laboratory of Mathematical Modeling and Scientific Computing, University of South China, Hengyang, 421001, China \\
$^3$Department of Physics, Xinzhou Normal University, Xinzhou 034000, Shanxi, China\\
$^4$College of Physics and Engineering Technology, Chengdu Normal University, Chengdu, Sichuan 611130, China
}

\begin{abstract}
Compared with the results of Planck-only analyses, recent measurements from the Atacama Cosmology Telescope (ACT) indicate a preference for a slightly bluer scalar spectral index,  placing canonical inflationary models in General Relativity (GR) under mild pressure. We demonstrate that $f(T)$ gravity systematically accommodates these dataset-dependent preferences by suppressing the tensor-to-scalar ratio in monomial and hilltop potentials, and by shifting the spectral index of E-models toward the ACT-favored region. Incorporating Big Bang Nucleosynthesis bounds, we break the degeneracy between the inflationary $e$-folding number and the post-inflationary thermal history. A direct side-by-side comparison reveals that reconciling models such as the Starobinsky potential with ACT data in GR strictly necessitates a non-standard, stiff (kinetic-dominated) reheating phase. In contrast, torsional corrections in $f(T)$ gravity significantly enlarge the viable parameter space, relaxing these stringent phenomenological requirements and establishing a coherent framework that jointly constrains CMB observables and reheating dynamics.
\end{abstract}

\pacs{}

\maketitle
\section{Introduction}
\label{sec_in}
The inflationary paradigm stands as a cornerstone of modern cosmology, successfully addressing the foundational puzzles of the Big Bang model while providing a quantum origin for the primordial density perturbations that seeded all cosmic structure \cite{STAROBINSKY198099, PhysRevLett.48.1220, PhysRevD.23.347, LINDE1982389}. High-precision measurements of the cosmic microwave background (CMB), led by the Planck satellite, have stringently tested this paradigm, mapping out the properties of these primordial fluctuations through observables such as the scalar spectral index $n_s$ and tensor-to-scalar ratio $r$ \cite{akrami2020planck, PhysRevLett.127.151301}.

However, this era of precision has also revealed subtle dataset-dependent differences in the preferred values of the scalar spectral index.  Recent results from the Atacama Cosmology Telescope (ACT) Collaboration~\cite{louis2025atacamacosmologytelescopedr6, calabrese2025atacamacosmologytelescopedr6}, particularly when combined with Planck data and other cosmological probes such as DESI BAO measurements \cite{Adame_2025, Adame2_2025} and BICEP/Keck B-mode data \cite{PhysRevLett.127.151301}, indicate a preference for a slightly higher scalar spectral index ($n_s \approx 0.9743 \pm 0.0034$ for P-ACT-LB) compared with that indicated by Planck-only findings ($n_s \approx 0.965 \pm 0.004$). It is crucial to clarify that this discrepancy between ACT and Planck is currently a mild statistical preference rather than a robust tension comparable to the Hubble crisis. Nevertheless, even a modest shift in $n_s$ places significant phenomenological pressure on several canonical inflationary models, such as Starobinsky inflation, Higgs inflation, and broad classes of $\alpha$-attractors, which were previously in excellent agreement with Planck-only constraints \cite{lynker2025actimplicationshilltopinflation,pallis2025actinspiredkaehlerbasedinflationaryattractors,peng2025polynomialpotentialinflationlight,Pallis_2025,wolf2025inflationaryattractorsradiativecorrections,haque2025minimalplateauinflationlight,Gao_2025,wang2025observationalconstraintsinflationarydecoherence,kallosh2025simplescenarioact,gao2025nonminimalcouplinglightact,Yin:2025rrs,Yogesh:2025wak,Yi:2025dms,Maity:2025czp,Heidarian:2025drk,Pal:2025ewf,chen2025probingreheatingdecayingoscillatory,Ketov:2025cqg,German:2025ide}. 
This situation motivates further exploration of either more flexible inflationary constructions within General Relativity (GR) or modest extensions to the gravitational framework itself.

This latter possibility has fueled interest in modified gravity theories as a means to reconcile inflationary theory with observation. Among the various proposals, teleparallel gravity offers a particularly compelling alternative \cite{unzicker2005translationeinsteinsattemptunified, Ferraro_2007, Ferraro_2008,Bengochea_2009, PhysRevD.81.127301, Wu_2010, FU_2011, Cai:2011tc, Wu_20112,Cardone_2012,Geng_2012, LIU_2012, Li_2013, PhysRevD.88.103526, Nesseris_2013, Iorio_2012,capozziello2015transitionredshiftftcosmology,Nunes_2017,Capozziello_2017, Otalora_2017, Xu:2018npu, Cai_2016,Bahamonde_2023,Capozziello:2024lsz, Bamba_2017, Nunes_2016, kumar2025higgslikeinflationscalartorsionftphi}. In this approach, gravitation is formulated through spacetime torsion on a globally flat manifold rather than curvature on a Riemannian one. The simplest version, known as the Teleparallel Equivalent of General Relativity (TEGR), reproduces the same field equations as GR. By generalizing the Lagrangian density to a function \(f(T)\) of the torsion scalar \(T\), in analogy with \(f(R)\) gravity~\cite{Buchdahl:1970ldb}, one obtains a natural second-order extension that avoids the higher-derivative instabilities of many modified gravity models while offering a distinctive and predictive phenomenology~\cite{Wu_20102, Wu_2011, Wei_2011, Karami_2013, Karami_20132,Capozziello_2022, Capozziello_2023}. In inflationary scenarios, $f(T)$ modifications alter the relation between the inflaton potential and the Hubble rate,  which typically reduces the tensor-to-scalar ratio $r$ relative to the GR prediction, thereby modifying the inflationary predictions for $(n_s, r)$ relative to the GR case and allowing an exploration of how torsional effects shift the observable parameter space~\cite{Rezazadeh_2016,Rezazadeh_2017,El_Bourakadi_2022,Zhang:2024gte,Jawad:2022sbx}.

However, modifying the gravitational sector is only one part of the puzzle. Precise constraints on inflationary models also require a detailed understanding of the post-inflationary thermal history, specifically the reheating epoch \cite{Abbott:1982hn, Dolgov:1982th, Albrecht:1982mp}. The uncertainty in the duration of reheating and the effective equation of state during this phase introduces a degeneracy in the prediction of the inflationary $e$-folding number, which in turn affects the theoretical values of $n_s$ and $r$~\cite{PhysRevD.82.023511, PhysRevLett.114.081303, Cook_2015, PhysRevD.92.063506, PhysRevD.103.103540, PhysRevLett.113.041302, PhysRevD.104.103530, PhysRevD.93.083524, ZHANG2023101169, PhysRevD.95.103502, Goswami_2018, Maity_2019, DENG2022101135,Zhang:2025aak,Zhang:2025tpg, Mishra_2021, Gong_2015, PhysRevD.102.021301, Garcia:2023tkk, Germ_n_2023, ZHANG2024138765,odintsov2025actinflationinfluencereheating}. In light of the improved precision of recent CMB datasets, including ACT, relying solely on standard assumptions for the duration of inflation may obscure important degeneracies between inflationary dynamics and the post-inflationary thermal history. Therefore, a robust analysis must treat the reheating temperature and the equation of state parameter as integral components of the constraints. In a previous study \cite{Zhang:2024gte} explored  the $f(T)$ model with power-law inflation against Planck-era data, and recent studies have examined reheating constraints under ACT data \cite{Liu:2025qca, Drees:2025ngb, Zharov:2025evb, Haque:2025uis}, However, a unified analysis combining the specific torsional effects of $f(T)$ gravity with rigorous reheating constraints across different potential classes using the latest ACT datasets remains absent.

In this study, we addressed this gap by performing a comprehensive comparative study of three benchmark potential classes, including power-law monomials, hilltop models, and E-models, using the $f(T) = C T^{2\delta+1}$ parameterization and the latest ACT data. We aim to elucidate how the torsional parameter $\delta$ uniquely modulates the $(n_s, r)$ predictions for these different potential shapes in light of the ACT-preferred region of parameter space, while simultaneously enforcing consistency with the thermal history of the Universe through rigorous reheating constraints, such as the Big Bang Nucleosynthesis (BBN) temperature limit. 
Through this unified approach, we aimed to characterize how deviations from GR manifest under combined inflationary and reheating considerations.

The remainder of the paper is organized as follows. In Sec.~\ref{sec:theory}, we briefly review the basic framework of $f(T)$ gravity and its application to inflationary cosmology. Section~\ref{sec:potentials} presents the analysis of the three potential classes and their inflationary predictions against ACT data. Section~\ref{sec:reheating} details the reheating analysis, deriving the constraints on the reheating temperature and duration for each model. We summarize and discuss the implications in Sec.~\ref{sec:conclusions}. Throughout this study, we adopt the metric signature \((-,+,+,+)\), set \(c=\hbar=1\), and work with reduced Planck units \(M_{\rm Pl}\equiv (8\pi G)^{-1/2}=1\).

\section{Inflationary dynamics in $f(T)$ gravity}
\label{sec:theory}
We consider a canonical single-field inflationary scenario within the teleparallel gravity framework, in which gravitation is described by torsion rather than curvature. The gravitational sector is generalized via a function $f(T)$ of the torsion scalar $T$, and the inflaton $\phi$ is minimally coupled to the metric. The action is given by~\cite{Ferraro_2007,Ferraro_2008,PhysRevD.88.103526}
\begin{align}
S = \int e\, d^{4}x \left[ \frac{1}{2} f(T) - \frac{1}{2} g^{\mu\nu} \partial_{\mu}\phi \, \partial_{\nu}\phi - V(\phi) \right],
\label{act}
\end{align}
where $e \equiv \det(e^{a}{}_{\mu}) = \sqrt{-g}$ is the determinant of the vierbein $e^{a}{}_{\mu}$, related to the metric tensor via $g_{\mu\nu} = \eta_{ab} e^{a}{}_{\mu} e^{b}{}_{\nu}$. Here $V(\phi)$ denotes the inflaton potential.
In teleparallel gravity, the torsion scalar is defined by
\begin{align}
T = S_{\sigma}{}^{\mu\nu} T^{\sigma}{}_{\mu\nu},
\label{T}
\end{align}
where the torsion tensor $T^{\sigma}{}_{\mu\nu}$ and  superpotential $S_{\sigma}{}^{\mu\nu}$ are given by
\begin{align}
T^{\sigma}{}_{\mu\nu} &= e_{a}{}^{\sigma} \left( \partial_{\mu} e^{a}{}_{\nu} - \partial_{\nu} e^{a}{}_{\mu} \right), \label{TT} \\
S_{\sigma}{}^{\mu\nu} &= \frac12 \left( K^{\mu\nu}{}_{\sigma} + \delta^{\mu}_{\sigma} T^{\alpha\nu}{}_{\alpha} - \delta^{\nu}_{\sigma} T^{\alpha\mu}{}_{\alpha} \right), \label{S}
\end{align}
with the contorsion tensor defined as
\begin{align}
K^{\mu\nu}{}_{\sigma} = -\frac12 \left( T^{\mu\nu}{}_{\sigma} - T^{\nu\mu}{}_{\sigma} - T_{\sigma}{}^{\mu\nu} \right).
\label{K}
\end{align}

For a spatially flat, homogeneous, and isotropic background, the metric takes the standard Friedmann–Robertson–Walker (FRW) form
\begin{align}
ds^2 = -dt^2 + a^2(t)\, \delta_{ij} dx^i dx^j ,
\label{FRW}
\end{align}
where $a(t)$ is the scale factor. Varying the action expressed by Eq.\eqref{act} with respect to the vierbein yields the modified Friedmann equations~\cite{Wu_20102, Wu_2011, Wei_2011, Karami_2013, Karami_20132}:
\begin{align}
3H^2 &= \rho_{\phi} + \frac12 \left( 2T f_{,T} - T - f \right), \label{Frid1} \\
\dot{H} &= -\frac12 \left[ \rho_{\phi} + P_{\phi} + 4\dot{H} T f_{,TT} + 2\dot{H} f_{,T} - 2\dot{H} \right],
\label{Frid2}
\end{align}
where $H \equiv \dot{a}/a$ is the Hubble parameter, $f_{,T} \equiv df/dT$, $f_{,TT} \equiv d^2f/dT^2$, and the scalar field energy density and pressure are
\begin{align}
\rho_{\phi} = \frac12 \dot{\phi}^2 + V(\phi), \quad P_{\phi} = \frac12 \dot{\phi}^2 - V(\phi),
\end{align}
which satisfy the continuity equation $\dot{\rho}_{\phi} + 3H(\rho_{\phi} + P_{\phi}) = 0$. This leads to the Klein–Gordon equation
\begin{align}
\ddot{\phi} + 3H\dot{\phi} + V_{,\phi} = 0.
\label{evop}
\end{align}

In this study, we focused on the power-law form of $f(T)$:
\begin{align}
f(T) = C\, T^{2\delta+1}, \quad C \equiv \frac{1}{M^{4\delta}},
\label{ft}
\end{align}
where $\delta$ is a dimensionless constant and $M$ is a mass scale\footnote{In subsequent calculations, we will set $C=1$. In addition, this specific power-law form is chosen to ensure real-valued solutions for $T=-6H^2 < 0$.}. Using $T=-6H^2$, the modified Friedmann equations become
\begin{align}
3H^2 &= \frac12 \left[ \frac{\rho_{\phi}}{C\left( 2\delta + \frac12 \right)} \right]^{\frac{1}{2\delta+1}}, \label{Frid11} \\
\dot{H} &= -\frac{3H^2}{2(2\delta+1)} \left( 1 + \frac{P_{\phi}}{\rho_{\phi}} \right).
\end{align}
The limit $\delta \to 0$ recovers GR. 
For convenience, we define the Hubble slow-roll parameters
\begin{align}
\epsilon_1 &\equiv -\frac{\dot{H}}{H^2} = \frac{3}{2(1+2\delta)}\left( 1 + \frac{P_{\phi}}{\rho_{\phi}} \right), \label{ep1} \\
\epsilon_2 &\equiv \frac{\dot{\epsilon}_1}{H\epsilon_1}.
\label{epi}
\end{align}
Inflation occurs when $\epsilon_1 \ll 1$ and $|\epsilon_2| \ll 1$, and ends when $\epsilon_1 \simeq 1$. From Eq.~\eqref{ep1}, using $P_\phi/\rho_\phi \le 1$, the maximal value of the first slow-roll parameter is $\epsilon_{1,\max} = {3}/({1+2\delta})$.
Requiring $\epsilon_{1,\max} \ge 1$ so that inflation can end gives $\delta \le 1$. Under slow-roll conditions ($\dot{\phi}^2 \ll V$, $|\ddot{\phi}| \ll |3H\dot{\phi}|$), one finds $\rho_{\phi} \simeq V(\phi)$ and $3H\dot{\phi} \simeq -V_{,\phi}$, leading to
\begin{align}
\epsilon_1 &\simeq \frac{C_1 V_{,\phi}^2}{2(1+2\delta)\, V^{\frac{2(1+\delta)}{1+2\delta}}}\label{sl1}, \\
\epsilon_2 &\simeq 2\epsilon_1 \left[ 2(1+\delta) - \frac{ 2(1+2\delta) V V_{,\phi\phi} }{ V_{,\phi}^2 } \right]\label{sl2},
\end{align}
where $C_1\equiv\left[ 4^{\delta} C (1+4\delta) \right]^{\frac{1}{1+2\delta}}$.
The number of $e$-folds is given by  
\begin{align}\label{Ns}
N_{\ast}\equiv\int_{t_{\ast}}^{t_{\mathrm{end}}} H\,dt
\simeq \int_{\phi_{\ast}}^{\phi_{\mathrm{end}}} 
-\frac{V^{\tfrac{1}{1+2\delta}}}{C_1\,V_{,\phi}}\, d\phi,
\end{align}
\noindent
between the horizon exit of the pivot scale 
$k_\ast $ (denoted by ``$\ast$'') 
and the end of inflation,
where $\phi_{\mathrm{end}}$ is determined by the condition $\epsilon_1(\phi_{\mathrm{end}}) = 1$.

The primordial scalar power spectrum in $f(T)$ gravity is~\cite{Cai:2011tc, Rezazadeh_2016, Rezazadeh_2017}
\begin{align}\label{ps}
P_s \simeq \frac{H^2}{8\pi^2 c_s^3 \epsilon_1} \bigg|_{c_s k = aH},
\end{align}
where the sound speed is
\begin{align}
c_s^2 = \frac{f_{,T}}{f_{,T} - 12H^2 f_{,TT}} = \frac{1}{1+4\delta}.
\label{cs}
\end{align}
From Eq.~\eqref{cs}, the positivity of $c_s^2$ demands $\delta > -1/4$, while the subluminal condition $c_s^2 \le 1$ enforces $\delta \ge 0$.  
Together with the requirement $\delta \le 1$ from $\epsilon_1$, this yields the viable range $0 \le \delta \le 1$ adopted hereafter.
Concerning the scalar amplitude, the recent P--ACT--LB analysis introduced the constrain~\cite{Adame2_2025}:
\begin{equation}\label{As}
\ln(10^{10}P_s(k_*)) = 3.060^{+0.011}_{-0.012},
\end{equation}
and at the pivot scale $k_*$,
the spectral index of the curvature perturbations $n_s$ and  tensor-to-scalar ratio $r$ can be expressed, respectively,  as~\cite{Rezazadeh_2017}
\begin{align}	n_s&\simeq1-2\epsilon_1(\phi_*)-\epsilon_2(\phi_*),\label{ns} \\
	r&\simeq 16c_s^3\epsilon_1(\phi_*)\label{r} .
\end{align}
It is important to clarify the physical role of the torsional parameter $\delta$. In $f(T)$ gravity, $\delta$ modifies the way in which the total energy density sources the cosmic expansion. The modified Friedmann relation can be expressed schematically as $H^2 \propto \rho^{\frac{1}{2\delta+1}},$ which differs fundamentally from the GR relation $H^2 \propto \rho$. Consequently, the Hubble slow-roll parameter $\epsilon_1 \equiv -\dot H/H^2$ is no longer determined solely by the inflaton potential, but is directly influenced by the torsional structure of the gravitational sector. Therefore, the resulting shifts in $(n_s, r)$  arise from genuine modifications of the background dynamics, rather than from an effective reparameterization of the inflationary model.
\section{Inflationary Potentials and Observational Constraints}
\label{sec:potentials}

In this section, we introduce the inflationary potentials considered in this study, which serve as representative models within the $f(T)$ gravity framework. For each potential, we present its theoretical form and compute the relevant inflationary observables, followed by a comparison with the latest ACT data.

\subsection{Power-law Potential}\label{Power}
\begin{figure}[H]
    \centering
    \includegraphics[width=0.7\textwidth]{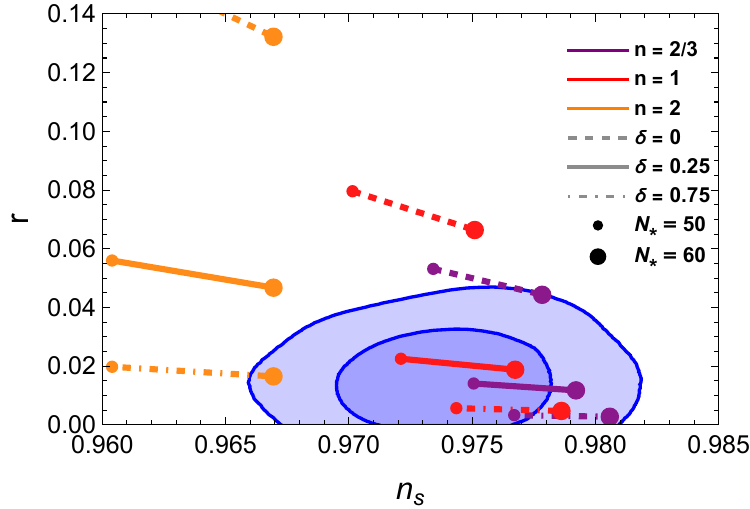}
    \caption{Inflationary predictions for the power-law potential in $f(T)$ gravity. The curves with different colors and line styles show the values of $n_s$ and $r$ as $n$ and $\delta$ vary. Each curve corresponds to a variation of the number of $e$-folds $N_*$ from $50$ to $60$.  The dark and light  blue shaded regions represent the $1\sigma$ and $2\sigma$ confidence intervals from the combined P-ACT-LB-BK$18$ data~\cite{Adame_2025, Adame2_2025}. }
    \label{fig1}
\end{figure}

We consider the potential form of the power-law model as follows~\cite{Linde:1983gd}:
\begin{equation}
V(\phi) = V_0 \phi^n,
\label{eq:powerlaw_potential}
\end{equation}
where $V_0$ is the normalization constant, and $n$ is the power index.
Substituting into Eq.~\eqref{sl1}, the first slow-roll parameter reads
\begin{equation}
\epsilon_1(\phi_*) \simeq  \frac{C_1 n^2 V_0^{\frac{2\delta}{1+2\delta}}}{2(1+2\delta) } \phi_*^{\frac{2n\delta}{1+2\delta}-2}.
\label{eq:epsilon1_powerlaw}
\end{equation}
Similarly, the second slow-roll parameter becomes
\begin{equation}
\epsilon_2(\phi_*) \simeq 2\epsilon_1(\phi_*) \left[ 2(1+\delta) - \frac{2(1+2\delta)(n-1)}{n} \right].
\label{eq:epsilon2_powerlaw}
\end{equation}
Using the slow-roll approximation and taking the limit $\phi_* \gg \phi_\mathrm{end}$, Eq.~\eqref{Ns} can be approximated by
\begin{equation}
N_* \simeq \frac{(1+2\delta)\,V_0^{-\frac{2\delta}{1+2\delta}}\,
      \phi_*^{\,2-\frac{2n\delta}{1+2\delta}}}
     {2\,n\,C_1\,[1+(2-n)\delta]}.
\end{equation}
From the above expressions, $n_s$ and $r$ follow straightforwardly via Eqs.~\eqref{ns} and \eqref{r} evaluated  as
\begin{align}	n_s&\simeq1-\frac{n}{2N_*\left[1+\delta\left(2-n \right) \right] }-\frac{1}{N_*},\label{ns2} \\
	r&\simeq\frac{4n}{N_*\left[1+\delta\left(2-n \right) \right]}\left( \frac{1}{1+4\delta}\right)^\frac{3}{2}\label{r2} .
\end{align}
Compared with GR ($\delta=0$), increasing $\delta$ raises $n_s$ for $n<2$, leaves it unchanged for $n=2$, and systematically suppresses the tensor-to-scalar ratio $r$ in all cases. Based on the above two equations, we can approximately obtain the spectral index and tensor-to-scalar ratio of the power-law potential for different values of $\delta$ at $N_* = 50$ and $60$,  presented in Tab.~\ref{TABLE 1}. In addition, we also present the numerical calculation results in Fig.~\ref{fig1} and compare them with the latest P-ACT-LB-BK$18$ observational data. From the figure, we conclude that the power-law potentials $n=2/3$, $1$, and $2$, which were originally excluded by the ACT data, can be consistent with observations under the influence of $f(T)$ gravity. Moreover, the greater the deviation of $f(T)$ gravity from GR, the better the consistency with observations. When $n=2$, $f(T)$ gravity deviates significantly ($\delta=0.75$) from GR and can agree with the P-ACT-LB-BK$18$ data at the $2\sigma$ level, supporting a larger $N_*$ (approaching $60$). For $n=2/3$ and $1$, it can even agree with observations at the $1\sigma$ level. While larger $\delta$ significantly suppresses $r$ for monomial potentials, allowing them to re-enter the viable region, the required $N_*$ values are sensitive to the post-inflationary evolution. We  rigorously constrain this dependency in Sec.~\ref{sec:reheating}.

\begin{table}[h]
\centering
\caption{\label{TABLE 1}Predicted $(n_s,\,r)$ for $V\propto\phi^{\,n}$ with coupling $\delta$ at $N_*=50, 60$.}
\setlength{\tabcolsep}{4pt}
\begin{tabular}{c|cc|cc|cc}
\hline\hline
$\delta$ & \multicolumn{2}{c|}{$n=2/3$} & \multicolumn{2}{c|}{$n=1$} & \multicolumn{2}{c}{$n=2$} \\
\hline
        & $N_*=50$      & $N_*=60$      & $N_*=50$      & $N_*=60$      & $N_*=50$      & $N_*=60$      \\
\hline
0       & (0.973,0.053) & (0.978,0.044) & (0.970,0.080) & (0.975,0.067) & (0.960,0.160) & (0.967,0.133) \\
0.25    & (0.975,0.014) & (0.979,0.012) & (0.972,0.023) & (0.977,0.019) & (0.960,0.057) & (0.967,0.047) \\
0.75    & (0.977,0.0033)& (0.981,0.0028)& (0.974,0.0057)& (0.979,0.0048)& (0.960,0.020) & (0.967,0.017) \\
\hline\hline
\end{tabular}
\end{table}

\subsection{Hilltop Inflation}
We consider the potential form of the hilltop model as follows~\cite{Boubekeur_2005}:
\begin{align}
V(\phi)=V_0\!\left[1- \left( \frac{\phi}{\mu} \right)^{p}\right],
\end{align}
where $p$ is a positive integer exponent, and $\mu$ sets the field scale.  For notational convenience, we also introduce the dimensionless field
\begin{align}
x\equiv \frac{\phi}{\mu}\in(0,1),
\end{align}
which will only be used to streamline intermediate expressions; all physical results are ultimately stated in terms of the original parameters. Substituting into Eqs.~\eqref{sl1} and \eqref{sl2} yields
\begin{align}
\epsilon_1(x_*) &\simeq \frac{C_1\,p^2}{2(1+2\delta)}\,\mu^{-2}\,V_0^{\frac{2\delta}{1+2\delta}}\,
x_*^{2p-2}\,\Big(1-x_*^{p}\Big)^{-\frac{2(1+\delta)}{1+2\delta}},
\label{eq:eps1_ht}\\[2pt]
\epsilon_2(x_*) &\simeq \frac{4\epsilon_1(x_*)}{p}\left\{(p-1)(1+2\delta)\,x_*^{-p}+\big[1-(p-2)\delta\big]\right\}.
\label{eq:eps2_ht}
\end{align}
The number of $e$-folds reads
\begin{align}
N_* \simeq \frac{\mu^2}{C_1\,p\,V_0^{\frac{2\delta}{1+2\delta}}}
\int_{x_*}^{x_{\rm end}}
x^{1-p}\,\big(1-x^p\big)^{\frac{1}{1+2\delta}}\,dx .
\label{eq:N_ht}
\end{align}
Using Eqs.~\eqref{ps} and \eqref{eq:eps1_ht}, together with the $f(T)$-modified Friedmann relation in Eq.\eqref{Frid11}, and working within the slow-roll regime where $\rho_\phi \simeq V$,
the scalar amplitude can be expressed so that the $V_0$-dependence is explicit:
\begin{align}
P_s
= \mathcal{K}(\delta)\;
\mu^2\,x_*^{\,2-2p}\,
(1-x_*^p)^{\frac{3+2\delta}{1+2\delta}}\;
V_0^{\frac{1-2\delta}{1+2\delta}},
\qquad
\mathcal{K}(\delta)\equiv\frac{1+2\delta}{12\pi^2\,C_1^2\,c_s^3\,p^2}.
\label{eq:As_powerlaw}
\end{align}

\subsubsection{Analytic Limits}
We consider two complementary limits of the hilltop potential, corresponding to the small-field regime near the maximum and the vicinity of the edge of the potential. 
\paragraph*{(i) Small-field regime, $x_*\ll1$ (near the hilltop).}
This regime typically arises when $\mu\ll1$, given that obtaining sufficient $e$-folds requires the inflaton to start very close to the hilltop. 
For $x\ll1$, one may expand $(1-x^p)^\alpha\simeq 1+\mathcal{O}(x^p)$. 
To leading order, Eq.\eqref{eq:N_ht} gives
\begin{align}
N_* &\simeq \frac{\mu^2}{C_1\,p\,(p-2)\,V_0^{\frac{2\delta}{1+2\delta}}}\,x_*^{\,2-p}\qquad (p>2),\label{Nss}
\end{align}
where the contribution of the upper limit $x_{\rm end}$ has been neglected\footnote{For $p>2$ and $x_*\ll1$, the relative correction from including $x_{\rm end}$ scales as $(x_{\rm end}/x_*)^{2-p}\ll1$, and is numerically subleading across the parameter space considered. In addition, the restriction $p>2$ ensures that the $e$-fold integral converges as $x\to0$, whereas for $p=2$ the result would become logarithmic rather than power-law.}.
Eliminating \(x_*\) yields the compact relations
\begin{align}
\epsilon_{1} (N_*)
&\simeq \frac{1}{2(1+2\delta)}\,
\frac{p^{\frac{2}{2-p}}}{(p-2)^{\frac{2(p-1)}{p-2}}}\,
\frac{\mu^{\frac{2p}{p-2}}}{\big(C_1\,V_0^{\frac{2\delta}{1+2\delta}}\big)^{\frac{p}{p-2}}}\;
N_*^{-\frac{2(p-1)}{p-2}},\label{eq:eps1_scaling_general}\\
\epsilon_{2} (N_*)
&\simeq \frac{2(p-1)}{(p-2)}\,\frac{1}{N_*}
\label{eq:eta_attractor}.
\end{align}
Combining Eq. \eqref{Nss} with Eq. \eqref{eq:As_powerlaw} eliminates $V_0$ and yields a closed form for $\epsilon_{1}$:
\begin{align}
\epsilon_{1}(N_*) 
\simeq
\frac{
\mu^{\tfrac{2p}{p+2p\delta-2}} \,
C_1^{\tfrac{p(2\delta-1)}{p+2p\delta-2}} \,
p^{\tfrac{4p\delta-2}{p+2p\delta-2}}\,\left[
P_s^{-1} \, \mathcal{K}(\delta)\right]^{\tfrac{2p\delta}{ p + 2p\delta-2}}
}{
2(1 + 2\delta)\,
(p-2)^{\tfrac{2(p-1)}{p + 2p\delta-2}}
}
\;
N_*^{-\tfrac{2(p-1)}{p + 2p\delta-2}}.
\label{eq:eps1}
\end{align}
Therefore, to leading order in the slow-roll era, the spectral index and  tensor-to-scalar ratio follow from Eqs.~\eqref{ns} and \eqref{r}, yielding
\begin{align}
n_s 
\simeq 1-\frac{2(p-1)}{(p-2)}\,\frac{1}{N_*}\,
\label{eq:ns_attractor}
\end{align}
and
\begin{align}
r \simeq \frac{8
\mu^{\tfrac{2p}{p+2p\delta-2}} \,
C_1^{\tfrac{p(2\delta-1)}{p+2p\delta-2}} \,
p^{\tfrac{4p\delta-2}{p+2p\delta-2}}\,\left[
P_s^{-1} \, \mathcal{K}(\delta)\right]^{\tfrac{2p\delta}{ p + 2p\delta-2}}
}{
(1 + 2\delta)\,\,(1+4\delta)^{3/2}\,
(p-2)^{\tfrac{2(p-1)}{p + 2p\delta-2}}
}
\;
N_*^{-\tfrac{2(p-1)}{p + 2p\delta-2}}.
\label{eq:r_attractor}
\end{align}
 From Eq.~\eqref{eq:ns_attractor}, the spectral tilt exhibits a universal hilltop attractor that is independent of both $\mu$ and $\delta$. In contrast, Eq.~\eqref{eq:r_attractor} shows that the tensor-to-scalar ratio depends explicitly on $p$, $\mu$, $\delta$, and $N_*$, reducing in the GR limit to the familiar scaling $r\simeq 8\,p^{\frac{2}{2-p}}\mu^{\frac{2p}{p-2}}\!\left[N_*(p-2)\right]^{\frac{2(p-1)}{2-p}}$.
As an illustration, for $p=4$, $\mu=10^{-4}$, and $\delta=0.25$, we obtain $n_s = 0.94$ and $r = 1.37\times10^{-8}$ for $N_*=50$, while for $N_*=60$ the results are $n_s = 0.95$ and $r = 1.04\times10^{-8}$. These analytic predictions are in good agreement with our numerical calculations.

\paragraph*{(ii) Vicinity of the edge, $x_*\to 1^-$.}
In the opposite regime, typically realized for $\mu\gg1$, inflation takes place near the boundary of the potential. 
We can write $x=1-\Delta$ with $0<\Delta\ll1$. A Taylor expansion gives
\begin{align}
V(\phi)=V_0\!\left[1-(1-\Delta)^p\right]
=V_0\!\left[p\,\Delta+\mathcal{O}(\Delta^2)\right],\qquad 
\Delta=\frac{\mu-\phi}{\mu}.
\end{align}
To leading order, the potential is linear in the displacement from the edge:
\begin{align}
V(\phi)\simeq \frac{p\,V_0}{\mu}\,(\mu-\phi).
\label{eq:linear_approx}
\end{align}
 Accordingly, the predictions approach those of linear inflation, as discussed in Subsection~\ref{Power}. As an example, $\delta=0.25$ and $N_*=50$ gives $n_s\simeq 0.972$ and $r\simeq 0.023$, in agreement with the numerical solutions presented below.

In summary, in the hilltop regime $x_*\ll1$, the scalar tilt exhibits a universal attractor independent of $\mu$, $V_0$, and $\delta$, namely Eq.~\eqref{eq:ns_attractor}, while the tensor-to-scalar ratio is strongly suppressed  in $f(T)$ gravity. 
In the opposite vicinity $x_*\to1^-$, the potential is effectively linear in the displacement from the edge.  
Both limits reduce smoothly to their GR counterparts as $\delta\to0$.

\subsubsection{Numerical Predictions and Observational Constraints}
\begin{figure}[H]
    \centering
    \includegraphics[width=0.7\textwidth]{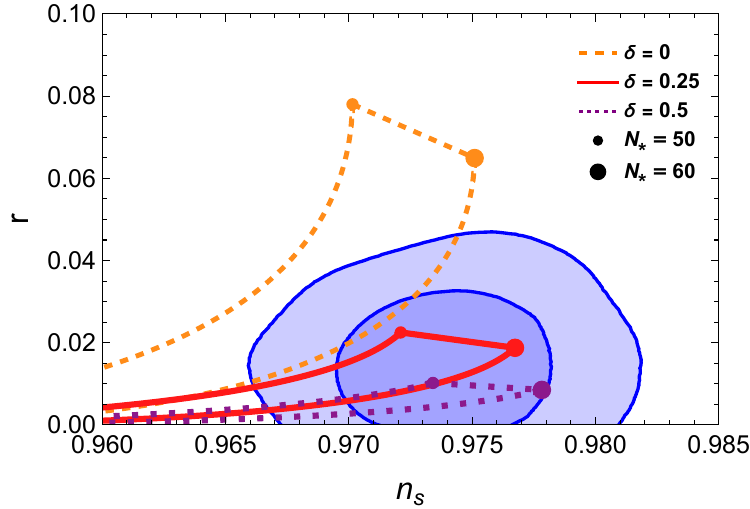}
    \caption{Inflationary predictions for the hilltop potential in $f(T)$ gravity. All shaded regions are the same as those in Fig.~\ref{fig1}. }
    \label{fig2}
\end{figure}

As a representative example, we set $p=4$ and computed $(n_s,r)$ numerically using Eqs.~\eqref{ns} and \eqref{r}. 
Fig.~\ref{fig2} displays the results for several values of the torsion parameter $\delta$, together with the latest P-ACT-LB-BK$18$ constraints in the $(n_s,r)$ plane. As demonstrated in the above limit analysis, the predicted values of $n_s$ and $r$ tend toward smaller values when $\mu \ll 1$. As $\mu$ increases, the predicted behavior gradually approaches the linear potential.
In the GR limit ($\delta=0$), the case $N_*=50$ is excluded, while for $N_*=60$ two viable intervals remain: $14.5 \lesssim \mu \lesssim 65.3$ at $2\sigma$ and $20.9 \lesssim \mu \lesssim 33.6$ at $1\sigma$. 
Turning on torsion-induced modifications shifts predictions toward larger $n_s$ and smaller $r$, thereby improving the agreement with data. 
For $\delta=0.25$, compatibility is already achieved at $N_*=50$ provided that $\mu \gtrsim 0.19$ ($2\sigma$) and $\mu \gtrsim 0.38$ ($1\sigma$); at $N_*=60$, the bounds relax to $\mu \gtrsim 0.09$ ($2\sigma$) and $\mu \gtrsim 0.14$ ($1\sigma$). 
A stronger suppression of $r$ occurs for $\delta=0.5$: at $N_*=50$, one finds that $\mu \gtrsim 0.002$ ($2\sigma$) and $\mu \gtrsim 0.0038$ ($1\sigma$); at $N_*=60$, the allowed region extends to $\mu \gtrsim 0.001$ ($2\sigma$) and $\mu \gtrsim 0.0015$ ($1\sigma$).

\subsection{E-model}
\label{sec:starobinsky}

We consider the E-model potential (a representative $\alpha$-attractor family) in the form~\cite{Kallosh_2013,Carrasco_2015}  
\begin{equation}
V(\phi)=V_0\bigl(1-e^{-A\phi}\bigr)^2,
\end{equation}
and introduce the variable $y\equiv e^{-A\phi}\in(0,1)$, such that the potential can be expressed as 
\begin{align}
V &= V_0(1-y)^2.
\end{align} 
In particular, setting $A=\sqrt{2/3}$ reproduces the classic Starobinsky potential (and is conformally equivalent to the Higgs inflation) \cite{STAROBINSKY198099}. While this model has long been favored by Planck data, the recent shift toward a slightly bluer spectral index indicated by ACT renders the standard GR prediction ($\delta=0$) increasingly disfavored within the ACT-preferred region, and may require comparatively specific reheating histories to remain within the $1\sigma$ contour \cite{Liu:2025qca}.
Using Eqs.~\eqref{sl1} and \eqref{sl2}, the Hubble slow-roll parameters can be expressed in closed form as functions of $y$: 
\begin{align}
\epsilon_1(y_*) &\simeq \frac{2A^2}{1+2\delta}\; C_1 \,V_0^{\frac{2\delta}{1+2\delta}}\; \frac{y_*^2}{(1-y_*)^{\tfrac{2}{1+2\delta}}}, 
\label{eq:eps1_staro_y} 
\\[2mm]
\epsilon_2(y_*) &\simeq2\,\epsilon_1(y_*)\left(\frac{1+2\delta}{y_*}-2\delta\right).
\label{eq:eps2_staro_y}
\end{align}
By changing the variable to $y=e^{-A\phi}$ (so that $d\phi=-dy/(A y)$), the number of $e$-folds can be expressed as 
\begin{align}
N_* &\simeq \frac{V_0^{-\frac{2\delta}{1+2\delta}}} {2A^2\,C_1} \int_{y_*}^{y_{\rm end}} \frac{(1-y)^{\frac{1-2\delta}{1+2\delta}}}{y^2}\,dy.
\label{eq:N_staro_int}
\end{align}

\subsubsection{Analytic limits}
\paragraph{Plateau (large-field) limit: $y_*\ll1$.}
the horizon exit on the plateau corresponds to $y_*= e^{-A\phi_*}\ll1$ (large $\phi_*$). In this regime, the integrand of Eq. \eqref{eq:N_staro_int} is dominated by $y_*$, and one may set $V\simeq V_0$ in the integrand to leading order.
 Expanding to lowest order in $y_*$, one  obtains the number of $e$-folds as
\begin{align}
N_* &\simeq \frac{1}{2A^2C_1}\,V_0^{-\frac{2\delta}{1+2\delta}}\,\frac{1}{y_*}.
\label{yN_plateau}
\end{align}
Substituting into Eqs. \eqref{eq:eps1_staro_y} and \eqref{eq:eps2_staro_y} yields
\begin{align}\label{ssl1}
\epsilon_{1}(N_*) &\simeq \frac{1}{2(1+2\delta)A^2C_1}\,V_0^{-\frac{2\delta}{1+2\delta}}\,\frac{1}{N_*^{2}},
\\
\epsilon_{2}(N_*) &\simeq \frac{2}{N_*}+\mathcal{O}(N_*^{-2}).
\end{align}
To eliminate \(V_0\), we can express $\epsilon_1$ by the observed amplitude to leading order from Eq.~\eqref{ps} as
\begin{equation}\label{As_plateau}
\epsilon_{1}(N_*) \simeq \frac{1}{2(1+2\delta)A^2C_1}\,\Bigg[\frac{12\pi^2c_s^3P_s}{\,(1+2\delta)A^2N_*^2}\Bigg]^{-\frac{2\delta}{1+2\delta}}\,\frac{1}{N_*^{2}} .
\end{equation}
 Therefore, the scalar tilt at leading order is the familiar plateau attractor.
\begin{align}
n_s \simeq 1-\frac{2}{N_*},
\label{eq:ns_plateau_clean}
\end{align}
while the tensor-to-scalar ratio is given by
\begin{align}
r &\simeq\frac{8\,c_s^3}{(1+2\delta)A^2 C_1}\;
\Bigg[\frac{12\pi^2c_s^3P_s}{\,(1+2\delta)A^2}\Bigg]^{-\frac{2\delta}{1+2\delta}}N_*^{-\tfrac{2}{1+2\delta}}.
\label{r_plateau_V0}
\end{align}
This expression reduces to the standard Starobinsky result $r\simeq 12/N_*^2$ when $\delta\to0$ and $A=\sqrt{2/3}$.   
For reference, when setting $N_*=50$ and $60$, the results obtained are $n_s\simeq 0.960,\quad r\simeq 4.8\times10^{-3}$ and $n_s\simeq 0.967,\quad r\simeq 3.3\times10^{-3}$ respectively,
which closely matches the widely known Starobinsky prediction.

\paragraph{Quadratic (small-field) limit: $A\phi_*\ll1$ (i.e. $y_*\simeq 1$).}
When the horizon exit occurs near the origin ($A\phi_*\ll1$), we may expand the exponential in the potential as
\begin{align}
V(\phi)=V_0\left(1-e^{-A\phi}\right)^2 
= V_0\left\{1-\left[1-A\phi+\mathcal{O}(\phi^2)\right]\right\}^2
\simeq V_0(A\phi)^2,
\end{align}
so that the inflationary dynamics is that of a quadratic potential.  The corresponding predictions have already been derived in  Subsection~\ref{Power}. When $N_*=50$ and $60$, the model predicts $n_s=0.960$ and $r=0.057$, and $n_s=0.967$ and $r=0.047$, respectively.

One can see that, at leading order, the scalar spectral index for the quadratic potential coincides with that obtained in the plateau regime \(n_s\simeq1-2/N_*\). Thus, \(n_s\) alone may not distinguish the two regimes at leading order.
 The difference is in the scaling of \(r\): plateau $\Rightarrow r\propto N_*^{-2/(1+2\delta)}$ (with additional $\delta$-suppression), while quadratic small-field $\Rightarrow r\propto N_*^{-1}$ (again multiplied by the sound-speed suppression $(1+4\delta)^{-3/2}$). This explains why, for moderate values of \(\delta\), the numerical results may show an $r$ larger than the naive plateau expectation but still much smaller than the GR quadratic value because of the \(c_s\) factor.

\subsubsection{Numerical Predictions and Observational Constraints}
\begin{figure}[H]
    \centering
    \includegraphics[width=0.7\textwidth]{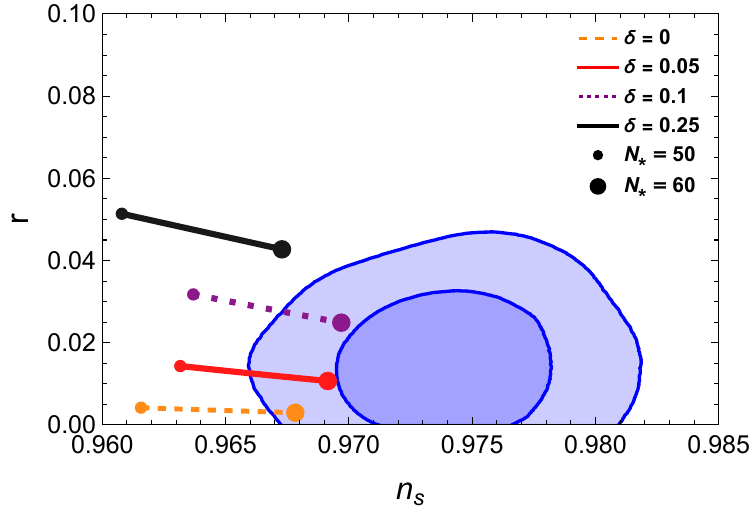}
    \caption{Inflationary predictions for the E-model with $A=\sqrt{2/3}$ in $f(T)$ gravity. Shaded regions correspond to the P-ACT-LB-BK$18$ constraints (color coding as in Fig.~\ref{fig1}).}
    \label{fig3}
\end{figure}

Fig.~\ref{fig3} displays our numerical results for the E-model with $A=\sqrt{2/3}$ in the $(n_s,r)$ plane, compared with the results obtained with the P-ACT-LB-BK$18$ constraints. For $\delta=0$, corresponding to the standard GR case, the Starobinsky model is found to lie almost entirely outside the allowed region. Once $\delta$ slightly departs from zero, i.e. when $f(T)$ gravity introduces small deviations from GR, the scalar spectral index $n_s$ and  tensor-to-scalar ratio $r$ increase. This trend substantially improves the agreement with observations: for $\delta=0.05$ and $\delta=0.1$, the predictions approach, and in some cases fall within, the $1\sigma$ confidence region of the P-ACT-LB-BK$18$ data. However, for larger deviations such as $\delta=0.25$, the dynamics of the Starobinsky potential resemble those of the quadratic potential, and the corresponding predictions shift away from the allowed region, eventually becoming disfavored by current CMB constraints.  This highlights a  mechanism distinct from the power-law and hilltop cases: here, $f(T)$ gravity acts to fine-tune the attractor prediction rather than just suppressing a large tensor signal.

\section{Reheating Analysis and Constraints}
\label{sec:reheating}
The epoch of reheating, bridging the end of inflation and the onset of the radiation-dominated era, is critical for determining the precise inflationary observables. The duration of this phase, characterized by the number of $e$-folds $N_{\mathrm{re}}$, and the effective equation of state $w_{\mathrm{re}}$, introduce a dependency in the relation between the pivot scale $k_*$ and total inflationary $e$-folds $N_*$. In this section, we quantify this connection within the $f(T)$ gravity framework to constrain the model parameters.

Based on the evolution of the comoving Hubble scale, the relation connecting the current CMB scale $k_*$ to the inflationary horizon crossing is given by:
\begin{align}
    \frac{c_s k_*}{a_0 H_0} = \frac{a_* H_*}{a_0 H_0} = \frac{a_*}{a_{\mathrm{end}}} \frac{a_{\mathrm{end}}}{a_{\mathrm{re}}} \frac{a_{\mathrm{re}}}{a_0} \frac{H_*}{H_0} = e^{-N_* - N_{\mathrm{re}}} \frac{a_{\mathrm{re}}}{a_0} \frac{H_*}{H_0},
    \label{eq:k_relation}
\end{align}
where the subscripts ``re'' and ``0'' denote the end of reheating and the present epoch, respectively. 
Assuming entropy conservation from the end of reheating to the present, the ratio of scale factors can be expressed in terms of temperatures:
\begin{align}
    \frac{a_{\mathrm{re}}}{a_0} = \left(\frac{43}{11 g_{\mathrm{s,re}}}\right)^{1/3} \frac{T_{\gamma}}{T_{\mathrm{re}}},
\end{align}
where $T_{\gamma}=2.725$K is the current CMB temperature, and $g_{\mathrm{s,re}}$ is the effective number of relativistic degrees of freedom for entropy at the end of reheating.
Combining these relations with the definition of the reheating equation of state, $\rho \propto a^{-3(1+w_{\mathrm{re}})}$, we derive the reheating temperature $T_{\mathrm{re}}$ as
\begin{align}
    T_{\mathrm{re}} = \left(\frac{43}{11 g_{\mathrm{s,re}}}\right)^{1/3} \frac{a_0 T_{\gamma} H_*}{c_s k_*} e^{-N_*} e^{-N_{\mathrm{re}}}.
    \label{eq:Tre_def}
\end{align}
Furthermore, from the continuity of energy density at the end of reheating, $\rho_{\mathrm{re}} = \frac{\pi^2}{30} g_{\mathrm{re}} T_{\mathrm{re}}^4 = \rho_{\mathrm{end}} e^{-3N_{\mathrm{re}}(1+w_{\mathrm{re}})}$, we obtain a canonical relation linking the inflationary and reheating parameters~\cite{PhysRevLett.113.041302, Cook_2015, Zhang:2024gte}:
\begin{align}
    (3w_{\mathrm{re}} - 1) N_{\mathrm{re}} = \ln \left( \frac{45 V_{\mathrm{end}}}{\pi^2 g_{\mathrm{re}}} \right) + \frac{1}{3} \ln \left( \frac{11 g_{\mathrm{re}}}{43} \right) + 4 \ln \left( \frac{c_s k_*}{a_0 T_{\gamma} H_*} \right) + 4 N_*.
    \label{eq:reheating_master}
\end{align}
Here, we have approximated $\rho_{\mathrm{end}} \simeq \frac{3}{2} V_{\mathrm{end}}(\phi_{\mathrm{end}})$ and assumed $g_{\mathrm{re}} = g_{\mathrm{s,re}} = 106.75$. This master equation allows us to solve for $N_{\mathrm{re}}$ and $T_{\mathrm{re}}$ as functions of the spectral index $n_s$ (via $N_*$) for a given potential and reheating scenario $w_{\mathrm{re}}$.
 Specifically, $H_*$ is related to the scalar amplitude $A_s$ via $H_*^2 \simeq 8\pi^2 c_s^3 \epsilon_1 A_s$, where $c_s$ depends on $\delta$.

We derive numerical constraints on the reheating epoch from Eq.~\eqref{eq:reheating_master} using the combined P-ACT-LB and BBN datasets. Our analysis covers three potential classes, starting with the power-law potential. Given that the reheating duration $N_{\text{re}}$ decouples from the spectral index for canonical reheating ($w_{\text{re}}=1/3$), we focused exclusively on the non-trivial cases where $w_{\text{re}} \neq 1/3$.

\subsection{Power-law Potential}
\label{subsec:res_power}
Based on the inflationary analysis in Sec.~\ref{sec:potentials}, only the fractional power-law models with $n=2/3$ and  linear potential $n=1$ remain viable within the $f(T)$ framework, specifically for non-vanishing torsional parameters $\delta=0.25$ and $0.75$. The quadratic model ($n=2$) requires an excessively large $\delta$ to marginally fit the data and is thus excluded from this detailed reheating analysis. We focused exclusively on the non-trivial cases where $w_{\mathrm{re}} \neq 1/3$. It is important to note that for the power-law potential in the GR limit ($\delta=0$), the tensor-to-scalar ratio remains well above the observational upper bound for any reasonable duration of inflation (e.g., $N_* \in [50, 60]$). Given that reheating dynamics cannot sufficiently suppress $r$ without requiring physically implausible e-folding numbers, we do not present the GR constraints for this case, as the model is already excluded by the B-mode polarization data regardless of the thermal history.

\begin{figure}[htbp]
    \centering
    \begin{tabular}{cc}
        \includegraphics[width=0.5\textwidth]{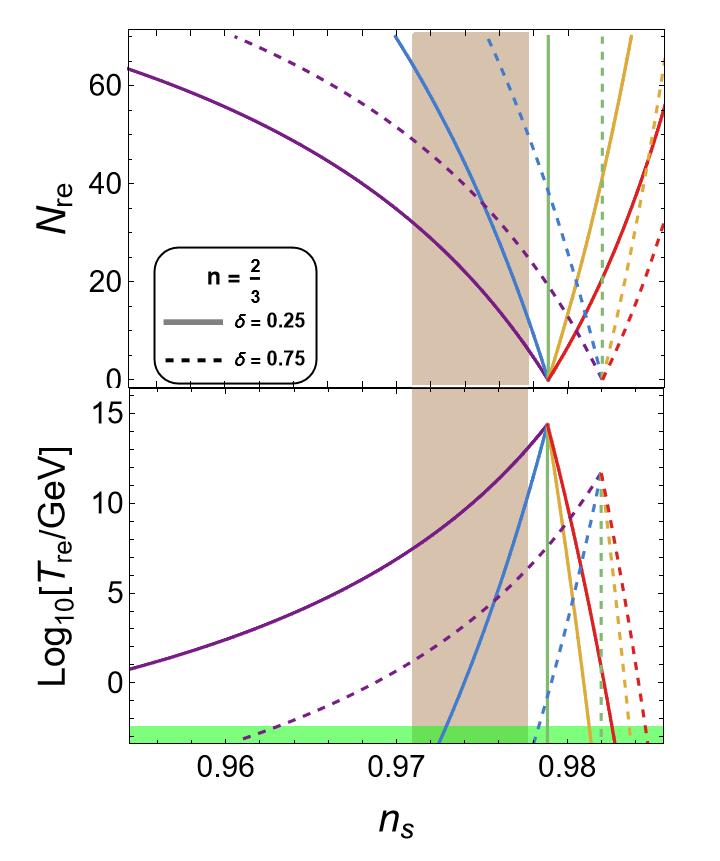} &
        \includegraphics[width=0.5\textwidth]{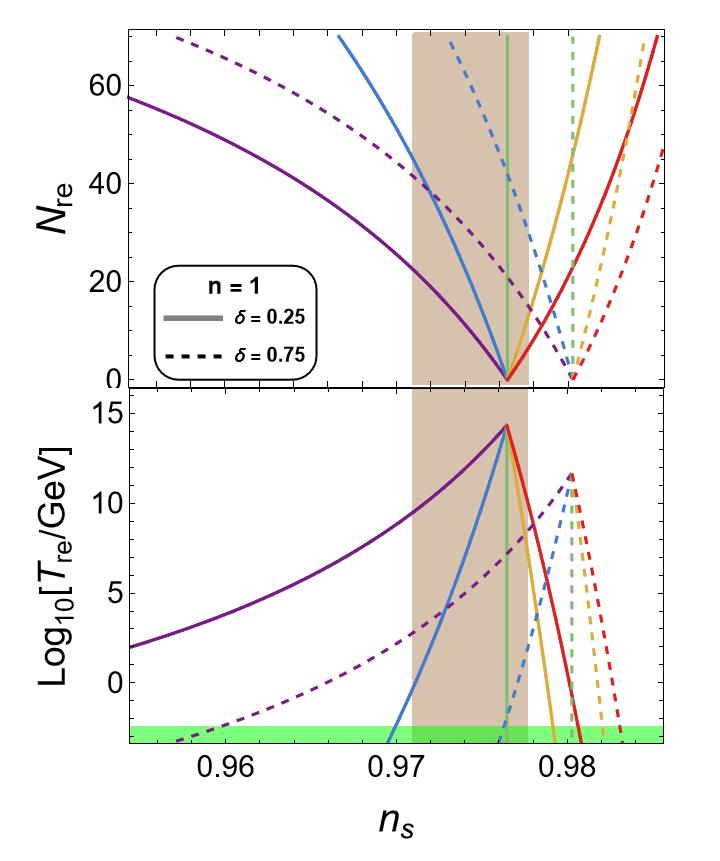} \\
    \end{tabular}
    \caption{
$N_{\text{re}}$ and $\log_{10}[T_{\text{re}}/\mathrm{GeV}]$ as functions of $n_s$ for the power-law potential with various values of $w_{\text{re}}$.  The colored curves represent different values of the equation-of-state parameter during reheating $w_{\text{re}}$: purple ($-1/3$), blue ($0$), green ($1/3$), orange ($2/3$), and red ($1$). The brown shaded region corresponds to the $1\sigma$ constraints from P-ACT-LB~\cite{louis2025atacamacosmologytelescopedr6}. The green band indicates the BBN-excluded region ($T_{\text{re}} < 4$ MeV) \cite{Kawasaki_1999, Kawasaki_2000, Hannestad_2004, Hasegawa_2019}.
}
    \label{fig4}
\end{figure}

Fig.~\ref{fig4} shows the number of $e$-folds $N_{\mathrm{re}}$ and the reheating temperature $T_{\mathrm{re}}$ as functions of the spectral index $n_s$. 
For the $n=2/3$ potential, which is favored by the data, the following results are obtained:
\begin{itemize}
    \item In the case of $\delta=0.25$, consistency with the P-ACT-LB constraints is achieved only for softer equations of state, specifically $w_{\mathrm{re}} = -1/3$ and $0$. The BBN temperature requirement further tightens the bounds. Solving Eq.~\eqref{eq:reheating_master} yields  precise inflationary $e$-folding intervals: $43.0 \le N_* \le 55.8$ for $w_{\mathrm{re}} = -1/3$, and $46.1 \le N_* \le 55.9$ for $w_{\mathrm{re}} = 0$.
    \item For the larger torsional correction $\delta=0.75$, the constraints are more severe. Only the $w_{\mathrm{re}} = -1/3$ scenario remains compatible with the ACT data, requiring a relatively high number of inflationary $e$-folds, $56.5 \le N_* \le 60.5$.
\end{itemize}
For the $n=1$ (linear) potential,  the following results are obtained: 
\begin{itemize}
    \item With $\delta=0.25$, the model exhibits broad compatibility. Instantaneous reheating ($N_{\mathrm{re}}=0$) is permitted within the $1\sigma$ ACT region. A wide range of reheating equations of state ($w_{\mathrm{re}} = -1/3, 0, 2/3, 1$) are viable. The corresponding constraints on the inflationary duration are derived as: $47.8 \le N_* \le 59.3$ ($w_{\mathrm{re}}=-1/3$), $48.0 \le N_* \le 59.3$ ($w_{\mathrm{re}}=0$), $59.3 \le N_* \le 62.4$ ($w_{\mathrm{re}}=2/3$), and $59.3 \le N_* \le 62.5$ ($w_{\mathrm{re}}=1$). Notably, stiffer\footnote{We defer a detailed physical discussion of the stiff reheating solutions to the E-model subsection below, where the general interpretation of such phases is clarified.} equations of state ($w_{\mathrm{re}} > 1/3$) necessitate a longer period of inflation ($N_* \gtrsim 59$).
    \item With $\delta=0.75$, the viable space shrinks again to softer reheating scenarios ($w_{\mathrm{re}} = -1/3, 0$), yielding $N_*$ ranges of $[44.0, 57.4]$ and $[54.1, 57.5]$, respectively.
\end{itemize}

\subsection{Hilltop Inflation}
\begin{figure}[htbp]
    \centering
    \begin{tabular}{cc}
         \includegraphics[width=0.33\textwidth]{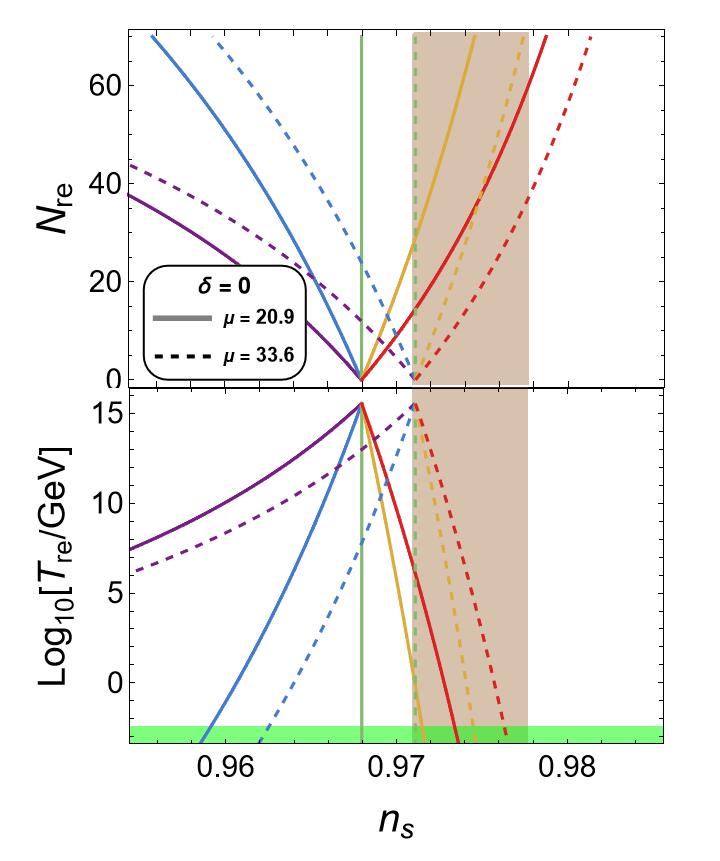}
        \includegraphics[width=0.33\textwidth]{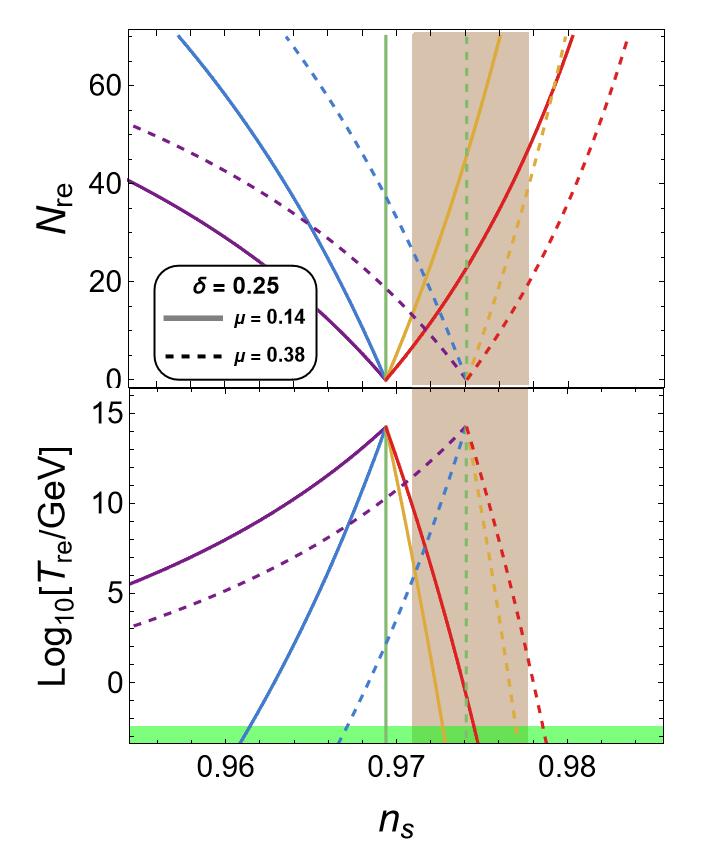} &
        \includegraphics[width=0.33\textwidth]{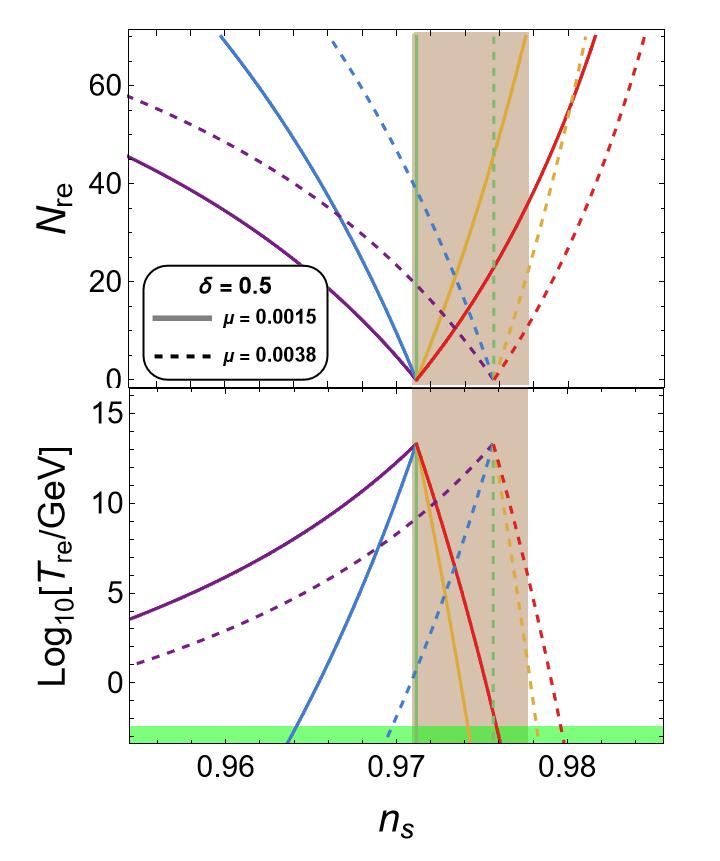} \\
    \end{tabular}
    \caption{Constraints on the reheating epoch for the hilltop potential. The plots show $N_{\mathrm{re}}$  and $\log_{10}[T_{\mathrm{re}}/\mathrm{GeV}]$ as functions of the spectral index $n_s$ for various values of $w_{\mathrm{re}}$. The color coding and shaded regions follow the conventions of Fig.~\ref{fig4}.}
    \label{fig5}
\end{figure}

Fig.~\ref{fig5} illustrates the reheating constraints for the hilltop potential, showing $N_{\mathrm{re}}$ and $T_{\mathrm{re}}$ as functions of $n_s$. To provide a direct comparison with GR, we included the GR predictions ($\delta=0$) alongside the $f(T)$ results ($\delta=0.25, 0.5$). For the GR case, we selected two representative field scales: $\mu=20.9$ and $\mu=33.6$. These values correspond to the lower and upper bounds of the parameter space where the GR prediction enters the $1\sigma$ region of the $(n_s, r)$ plane for $N_*=60$, as shown in Fig.~\ref{fig2}.

 In the GR limit, the smaller field scale $\mu=20.9$ is compatible with ACT data only for stiff reheating equations of state, $w_{\mathrm{re}} = 2/3$ and $w_{\mathrm{re}} = 1$. When combined with BBN bounds, these scenarios impose tight constraints on the inflationary duration: $63.0 \leq N_* \leq 64.4$ and $63.1 \leq N_* \leq 69.8$, respectively. For the larger field scale $\mu=33.6$, the model favors instantaneous reheating. The allowed e-folding intervals are derived as $56.4 \leq N_* \leq 64.6$ for $w_{\mathrm{re}} = 2/3$ and $56.4 \leq N_* \leq 70.1$ for $w_{\mathrm{re}} = 1$.

For the case of $\delta = 0.25$, we first consider $\mu = 0.14$ (corresponding to the $N_* = 60$ benchmark). Here, consistency with ACT data is found only for stiffer equations of state, specifically $w_{\mathrm{re}} = 2/3$ and $1$. The combined BBN and spectral index constraints restrict the e-folding number to $62.3 \leq N_* \leq 66.6$ and $62.3 \leq N_* \leq 71.6$, respectively. Conversely, for $\mu = 0.38$ (corresponding to the $N_* = 50$ benchmark), the scenario of instantaneous reheating becomes viable. The allowed ranges for $N_*$ were found to be $[52.4, 59.1]$ for $w_{\mathrm{re}} = -1/3$, $[52.5, 59.1]$ for $w_{\mathrm{re}} = 0$, $[59.1, 66.8]$ for $w_{\mathrm{re}} = 2/3$, and $[59.1, 68.9]$ for $w_{\mathrm{re}} = 1$.

In the case of $\delta = 0.5$, we set $\mu = 0.0015$ for the $N_* = 60$ benchmark. Instantaneous reheating is permitted, with valid solutions found for $w_{\mathrm{re}} = 2/3$ and $1$, yielding constraints of $61.9 \leq N_* \leq 69.1$ and $61.9 \leq N_* \leq 73.9$, respectively. Finally, for $\mu = 0.0038$ (the $N_* = 50$ benchmark), instantaneous reheating remains consistent with the data. The derived bounds on the e-folding number are $52.0 \leq N_* \leq 62.0$ ($w_{\mathrm{re}} = -1/3$), $51.9 \leq N_* \leq 62.0$ ($w_{\mathrm{re}} = 0$), $62.0 \leq N_* \leq 67.7$ ($w_{\mathrm{re}} = 2/3$), and $62.0 \leq N_* \leq 67.6$ ($w_{\mathrm{re}} = 1$).

\subsection{E-model}
Finally, we examine the reheating predictions for the E-model with $A=\sqrt{2/3}$ (Starobinsky potential). To explicitly quantify the impact of torsional modifications, we performed a side-by-side analysis of the standard GR limit ($\delta=0$) and the $f(T)$ scenarios ($\delta=0.05, 0.1$).

\begin{figure}[H]
    \centering
    \includegraphics[width=0.6\textwidth]{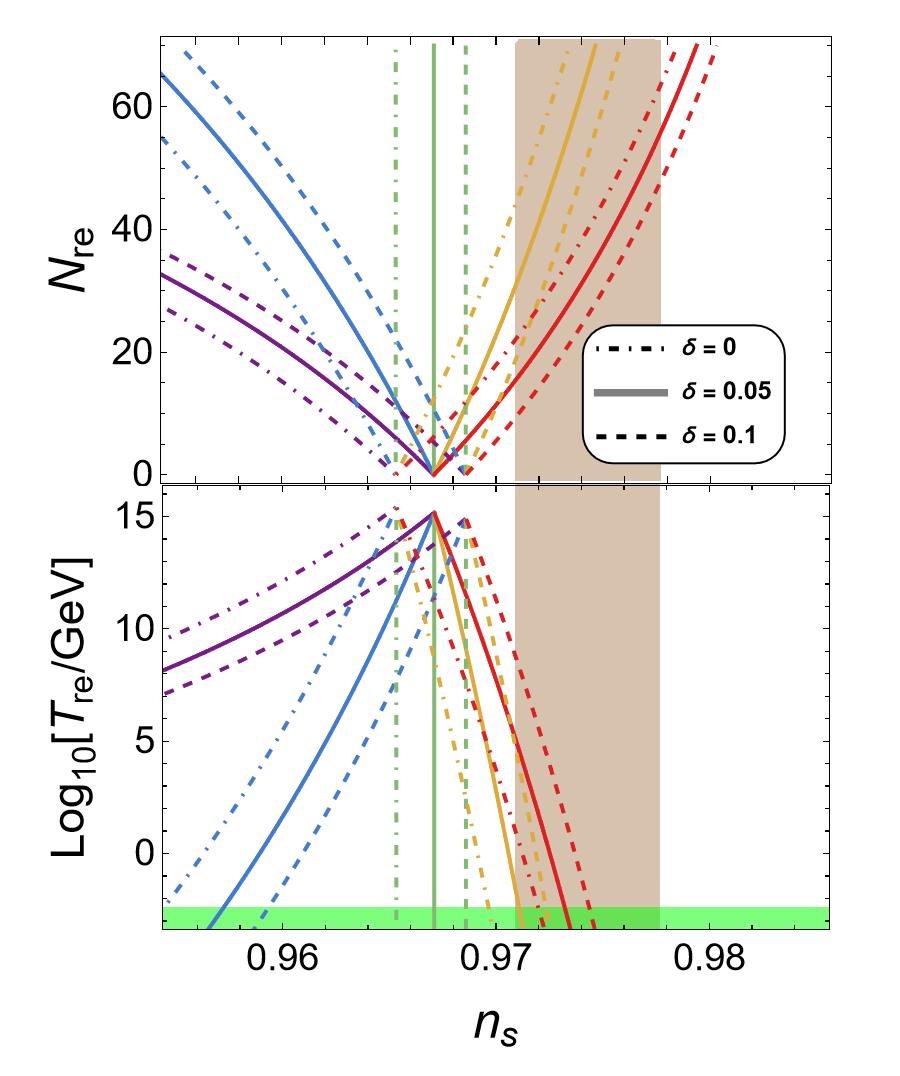}
    \caption{Constraints on the reheating epoch for the E-model with $A=\sqrt{2/3}$. The plots show $N_{\mathrm{re}}$ and $\log_{10}[T_{\mathrm{re}}/\mathrm{GeV}]$ as functions of the spectral index $n_s$ for various values of $w_{\mathrm{re}}$. The color coding and shaded regions follow the conventions of Fig.~\ref{fig4}.}
    \label{fig6}
\end{figure}

In Fig.~\ref{fig6} , the GR predictions barely intersect the ACT-favored region. Consistency is achieved exclusively for a stiff, kinetic-dominated equation of state ($w_{\mathrm{re}}=1$). Furthermore, this scenario demands a prolonged period of inflation, with the e-folding number restricted to a narrow range of $66.5 \leq N_* \leq 69.1$. This implies that within standard GR, the Starobinsky model can only be reconciled with ACT data by invoking a specific, non-standard reheating history dominated by the kinetic energy of the inflaton.

In contrast, $f(T)$ gravity significantly relaxes these constraints. For $\delta=0.05$ and $\delta=0.1$, consistency with the combined P-ACT-LB and BBN datasets is achieved for  less stiff equations of state, specifically $w_{\mathrm{re}} = 2/3$ and $w_{\mathrm{re}} = 1$. Consequently, we obtained the following constraints on the number of e-folds $N_*$: for the $\delta = 0.05$ case, the allowed ranges are restricted to $64.5 \leq N_* \leq 64.8$ ($w_{\mathrm{re}}=2/3$) and $64.5 \leq N_* \leq 70.2$ ($w_{\mathrm{re}}=1$). Similarly, for $\delta = 0.1$, the constraints correspond to $62.5 \leq N_* \leq 65.9$ and $62.5 \leq N_* \leq 71.2$ for $w_{\mathrm{re}} = 2/3$ and $w_{\mathrm{re}} = 1$, respectively. This comparison highlights that while GR requires a specific kination epoch to remain viable, weak torsional corrections naturally expand the allowed parameter space, accommodating a broader range of thermal histories.

Before concluding this section, we address two physical implications of our reheating analysis. The first concerns the frequent appearance of a stiff reheating phase ($w_{\mathrm{re}} \gtrsim 2/3$) in the viable regions of the Hilltop and E-models. While standard perturbative reheating typically leads to radiation-like ($w=1/3$) or matter-like ($w=0$) equations of state, a stiff phase (e.g., $w \simeq 1$) corresponds to a kination epoch where the kinetic energy of a scalar field temporarily dominates over its potential energy \cite{Spokoiny_1993,Joyce_1997, Dimopoulos_2018}. We stress that such a stiff phase is not assumed here as a generic prediction of specific particle-physics reheating models. Instead,  the parameter $w_{\mathrm{re}}$ serves as an effective, macroscopic description of the post-inflationary expansion history. The stiff solutions identified above should be interpreted as phenomenological allowances required by the CMB data, rather than as direct evidence for a specific microphysical mechanism.

The second issue relates to parameter correlations.  Although the BBN bounds effectively restrict the allowed range of $N_*$ and compress the viable interval of the torsional parameter $\delta$, a residual degeneracy persists among $\delta$, $w_{\mathrm{re}}$, and the inflaton potential parameters (such as $\mu$).  These potential parameters dictate the background evolution near the end of inflation, thereby determining the required reheating history for a given $\delta$.   Consequently, a stronger torsional correction (larger $\delta$) may necessitate a severely restricted range of $w_{\mathrm{re}}$ and potential parameters to maintain $n_s$ within the ACT-preferred region.  The emergence of viable solutions within narrow parameter windows (e.g., E-models requiring specific $w_{\mathrm{re}}$) should not be viewed as artificial fine-tuning, but rather as a reflection of the strong predictive power of the combined constraints from inflationary observables, ACT data, and the thermal history.

\section{Discussion and Conclusions}
\label{sec:conclusions}
Compared with the results of Planck-only analyses, high-precision data from ACT indicate a preference for a slightly bluer scalar spectral index placing mild pressure on several canonical inflationary models within GR. In this study, we demonstrated that teleparallel $f(T)$ gravity provides a robust mechanism to accommodate this dataset-dependent preference. By incorporating torsion-induced dynamics parameterized by $\delta$, we successfully reconciled the predictions of power-law, hilltop, and E-models with the combined P-ACT-LB-BK$18$ and BBN constraints. This reconciliation originates from a genuine modification of the background expansion, which alters the slow-roll hierarchy rather than merely reparameterizing the potentials. This dynamic breaks the degeneracy between the inflationary duration $N_*$ and the thermal history, pinpointing precise parameter intervals that often extend beyond the standard $N_* \in [50, 60]$ assumption.

Our systematic analysis yields distinct phenomenological signatures for each potential class under the combined constraints of CMB data and BBN:
\begin{itemize}
    \item \textbf{Power-law Potentials:} Torsional corrections effectively rescue the fractional ($n=2/3$) and linear ($n=1$) potentials that are otherwise disfavored in GR. Specifically, the $f(T)$ modification significantly suppresses the tensor-to-scalar ratio while shifting the spectral index toward the bluer values favored by ACT. The reheating constraints indicate a preference for softer equations of state: the $n=2/3$ case requires $w_{\mathrm{re}} \le 0$, while the linear model remains compatible with a broader range of thermal histories, including instantaneous reheating.

   \item \textbf{Hilltop Models:} In the GR limit, consistency with ACT data restricts the viable parameter space to relatively large field scales ($\mu \gtrsim 20$) and imposes stringent limits on the thermal history. In contrast, $f(T)$ gravity effectively suppresses the tensor-to-scalar ratio $r$, significantly broadening the viable parameter space to include much smaller field scales ($\mu \ll 1$) while maintaining the observationally preferred high $n_s$. For these torsion-rescued small-field scenarios, matching the data strictly requires a stiff, kinetic-dominated reheating phase ($w_{\mathrm{re}} \ge 2/3$) and a larger number of $e$-folds ($N_* \gtrsim 62$). Larger field scales, however, remain compatible with standard reheating evolution.

     \item \textbf{E-models (Starobinsky Potential):} Our explicit comparison reveals that in the standard GR limit, the Starobinsky potential is under significant pressure, compatible with ACT data only if one assumes an extreme kinetic-dominated reheating phase ($w_{\mathrm{re}}=1$) and a large number of $e$-folds ($N_* \gtrsim 66$). Introducing weak torsional corrections ($\delta \approx 0.05-0.1$) alleviates this severe restriction. By slightly enhancing the scalar spectral index relative to its GR prediction, $f(T)$ gravity allows consistency for a broader range of stiff equations of state (including $w_{\mathrm{re}}=2/3$) and widens the viable inflationary duration. This demonstrates that $f(T)$ gravity naturally restores the robustness of E-models against current observational bounds.
\end{itemize}

It is important to emphasize that the restoration of viability identified in this study is not an artifact of fine-tuning the specific ansatz $f(T)\sim T^{2\delta+1}$. The essential mechanism is the modification of the consistency relation between $H$ and $\rho$, which is a generic feature of power-law–type torsional corrections. While different functional forms of $f(T)$ would quantitatively shift the precise parameter boundaries, the qualitative alleviation of observational pressure persists for models introducing similar torsional corrections. Furthermore, while the combined BBN and CMB bounds restrict certain models to narrow parameter windows (such as the specific $w_{\mathrm{re}}$ required for E-models), this underscores the strong predictive power of the framework rather than indicating artificial fine-tuning.

In summary, $f(T)$ gravity serves as a distinct gravitational paradigm imposing specific, testable requirements on post-inflationary evolution. As observational precision continues to improve, the unique signatures identified in this study, including the requirement for non-standard stiff reheating epochs and larger $N_*$ values to match a bluer $n_s$, will serve as powerful discriminators between torsional modifications and standard Riemannian gravity.

\begin{acknowledgments}
This work was supported by the National Natural Science Foundation of China (Grant No. 12505075), the Natural Science Foundation of Sichuan Province, China (for Young Scientists) (Grant No. 2026NSFSC0808), and the Research Incentive Program for Doctors Joining Shanxi (Grant No. Z20240219).
\end{acknowledgments}

\bibliographystyle{apsrev4-1}
\bibliography{ref}
\end{document}